# Feedforward Compensation of Piezo Nonlinearity for High-Precision High-Speed Atomic Force Microscopy

Kenichi Umeda[1,2] and Noriyuki Kodera[1]

*[1] Nano Life Science Institute (WPI-NanoLSI),*

*Kanazawa University, Kakuma-machi, Kanazawa, Ishikawa, 920-1192, Japan.*

*[2] PRESTO/JST, 4-1-8 Honcho, Kawaguchi, Saitama 332-0012, Japan.*

Corresponding Authors

Dr. Kenichi Umeda (E-mail: umeda.k@staff.kanazawa-u.ac.jp)

Prof. Noriyuki Kodera (E-mail: nkodera@staff.kanazawa-u.ac.jp)

## Abstract

Atomic force microscopy (AFM) enables nanoscale characterization and has been widely applied to a broad range of systems. Over the past two decades, advances in high-speed AFM have enabled not only the imaging of static structures but also the direct observation of nanoscale dynamics in real time. However, because the tip or sample is typically scanned using piezoelectric actuators, nonlinearities in their response to the input signal can introduce image-scaling errors of up to 20–30%. Consequently, there is a strong demand for a method to correct piezoelectric nonlinearity that can reliably support quantitative dynamic structural measurements. Here, we propose a simple software-based feedforward method to generate scan waveforms that can be readily implemented. We identify four distinct sources of positioning error in piezo scanners and demonstrate that these errors can be compensated, achieving an order-of-magnitude improvement in positioning accuracy compared with uncompensated operation. Because the proposed method is software-based and requires no additional hardware, it preserves imaging speed and is well suited for high-speed AFM. It is also compatible with a wide range of AFM and other scanning probe systems.

# 1. Introduction

Atomic force microscopy (AFM) is a versatile nanoscale metrological technique for characterizing surface structures, by scanning an atomically sharpened tip over a sample[1-5]. Because AFM is widely used across diverse materials[4,6-12], accurate determination of structural dimensions is essential for reliable measurements. Over the past two decades, advances in high-speed AFM (HS-AFM) have enabled not only imaging of static structures but also direct observation of rapidly evolving samples, particularly biomolecules[9,12-24]. This capability allows transient structural changes to be quantified through measurements of binding sites, inter-domain distances and angles, step sizes, and molecular shape[14-30]. Because such structural information is closely linked to biological function, quantitative measurements compatible with HS-AFM are increasingly important for advancing scientific progress.

Piezoelectric scanners, devices that convert applied voltage into mechanical displacement[31], provide sub-nanometer positional resolution and can be driven at frequencies of tens of kilohertz or higher, making them indispensable for both conventional AFM and HS-AFM[10,12,32,33]. However, nonlinear piezo responses can cause deviations of up to 20−30% between the drive signal and the actual displacement, making compensation necessary[5,12,34].

Closed-loop control, typically implemented using capacitive displacement sensors in AFM instruments, measures the actual piezo displacement in real time and corrects the drive signal accordingly[32,35-37]. Although highly accurate, this approach suffers from sensor noise and limited high-frequency performance, making it unsuitable for HS-AFM. Charge control exploits the more linear relationship between electric charge and displacement in piezo elements and remains effective at high frequencies[38-42], but it suffers from low-frequency drift and leakage and cannot fully compensate for offset-voltage–dependent variations. Both methods also require additional hardware, increasing system cost.

Model-based feedforward compensation uses hysteresis models to correct nonlinearity without extra hardware[32,43-48]. However, these models often contain many parameters, making parameter identification difficult and limiting their adoption in AFM. Some commercial systems offer simplified hysteresis-compensation modes, but only in a restricted form[49,50]. Moreover, both charge-control and model-based approaches are affected by temperature changes, mechanical stress, or aging, so that calibration may not hold, with possible deviations of a few percent.

Post-image-processing–based hysteresis correction is also widely used[12,34,51-54], but it has several drawbacks: computational cost is incurred each time data are loaded, periodic or characteristic structures are required, and the backward scan is essential, making it incompatible with scanning strategies that omit the backward pass[55].

Regardless of the imaging method, regular calibration is essential for achieving high positional accuracy in AFM imaging. However, in many AFM applications, positional errors of 1–2% are often acceptable due to inherent uncertainties arising from thermal drift and tip convolution. Thus, practical factors such as ease of calibration and usability are important, especially in HS-AFM, where high bandwidth and low noise are prioritized. Furthermore, most previous studies have focused solely on hysteresis and dynamic compensation[12,13,47], whereas offset-voltage dependence and scan-size nonlinearity also exert significant influence on positioning errors.

Motivated by these considerations, we first identified four distinct sources of piezo nonlinearity and developed simple analytical compensation models for each, which are both practical and easy to implement. Although designed primarily to improve quantitative distance analysis in molecular AFM imaging, particularly HS-AFM, the method is also applicable to general AFM instruments. Requiring no specialized hardware or complex calibration, the proposed method improves positioning accuracy by more than an order of magnitude, making it useful for a wide range of AFM applications.

## 2. Background of piezo actuator

Ferroelectric materials exhibit spontaneous polarization, the direction of which can be controlled by an external electric field[56]. To reduce internal energy, the polarization is divided into multiple domains rather than being uniform throughout the material. Domain walls are pinned by defects, grain boundaries, and residual stresses, so not all domains readily switch under an average electric field[56,57]. Consequently, polarization increases gradually at low electric fields, rises steeply in the intermediate-field regime due to enhanced domain-wall motion, and saturates at high electric fields as domains become almost fully aligned and domain-wall motion is suppressed. These mechanisms give rise to pronounced hysteresis of ferroelectric materials (Fig. 1a).

Since unpoled ferroelectric materials exhibit a small macroscopic response due to randomly oriented domains, they are subjected to a poling process for use in practical actuators[31]. Figure 1b shows the experimental displacement–voltage characteristics of the typical HS-AFM X-piezo scanner; details are described in the following sections. As a result of poling, the relationship becomes nearly linear, especially at small voltages (red curve, Fig. 1b), where lattice strain dominates. However, at higher voltages, residual microscopic inhomogeneities still permit slight local domain-wall motion, leading to overall nonlinearity of the response (blue and green curves, Fig. 1b), which degrades positioning accuracy in AFM.

## 3. Methods

In this study, to evaluate piezo nonlinearities, we used a lab-built HS-AFM scanner that adopts an orthogonal XY stack-piezo configuration with a flexure-guided structure, in which the X- and Y-piezo actuators perform the fast and slow scans, respectively (Fig. 1c,d)[13]. Compared with the tube-type piezos commonly used in conventional AFM systems, stack-type piezo actuators offer higher resonance frequencies and relatively lower cross-talk between the XYZ piezos[5,10].

We evaluated the X-piezo scanner, a commercial piezo actuator (AE0203D08F, NEC Tokin) with a maximum drive voltage of 150 V and a resonance frequency of approximately 140 kHz. When installed in the scanner under preload, the resonance frequency decreased to approximately 40 kHz, and the piezoelectric coefficient was 12–20 nm/V.

A scan signal generated by custom-built control software (UMEX Sample-Scan HS-AFM) was output via a DA converter (±5 V range). This signal was amplified by a factor of ten using a piezo driver (HJPZ-0.1P×3, Matsusada Precision Inc., max. 100 V) and applied to the piezo scanner. For frequency-sweep measurements, a wideband piezo driver (M-2335, MESS-TEK Co., Ltd., maximum output 50 V) was used instead. Increasing the applied voltage extends the piezo (forward direction), whereas decreasing the voltage produces the backward direction.

For AFM imaging, we measured Annexin V two-dimensional (2D) crystal lattices formed on a lipid membrane spread on mica using a typical HS-AFM setup[13]. The 2D crystals were prepared as described previously[12,26,58]. For the Annexin imagings, a Fourier filter was applied slightly to enhance the periodic lattice structure. To characterize the hysteresis behavior, a grating sample (TGQ1, NT-MDT) was imaged using an ultra-wide scanner with a maximum scan range of approximately 30 μm (see Fig. 4b,d)[12].

Direct piezo displacement was measured using a Doppler laser interferometer (VibroOne, Polytec GmbH) or a heterodyne laser displacement sensor (ST-3761, IWATSU). Frequency sweeps were carried out using a digital lock-in amplifier (HF2LI, Zurich Instruments) operated by custom software.

## 4. Overview of piezo nonlinearity

Through a series of measurements, we classified the nonlinear piezo characteristics affecting AFM positioning accuracy into four major categories, shown in Table 1 and Fig. 1e:

**Table 1.** Major categories of piezo nonlinearity and their contributions to imaging error

| Type of nonlinearity | Imaging error | Contribution order |
| --- | --- | --- |
| Tip position (offset voltage) dependency | Max. 30% | 1 |
| Scan size nonlinearity | Max. ~25% (typically <15%) | 2 |
| Scan wave hysteresis | Max. 5−10% | 3 |
| Scan frequency | Max. ~1−2% | 4 |

In our measurements, these effects affected the scan size in the order described above. While some commercial instruments include functions to compensate for scan-wave hysteresis via analytical models[49,50], correcting all four components is uncommon. In the following sections, we describe the characteristics of each effect and the corresponding compensation based on analytical models.

## 5. Tip position (offset voltage) dependency

In AFM imaging, when the density of structures of interest, such as biomolecules, on the substrate is low, a search scan is required to locate the target structure. As illustrated in Fig. 2a, typical AFM control software allows the user to freely adjust the XY tip position. Some commercial AFM systems employ an additional positioner independent of the imaging scanner; however, this approach increases system size, limits high-speed operation, and raises overall cost. Therefore, many AFM systems, including our HS-AFM, achieve positioning by superimposing an offset voltage onto the scan voltage of the imaging scanner.

However, as described in this section, the offset voltage affects the AC piezoelectric coefficient because of piezoelectric nonlinearity. This effect represents the largest source of positioning error in biomolecular imaging. To evaluate its impact, we performed AFM imaging of an Annexin V crystal. Imaging was first performed near the center of the positioning range (XY offset voltage = 50 V). The image size was 200 × 200 nm, corresponding to ~13 V, or 13% of the maximum voltage. In Fig. 2b, clear hexagonal 2D lattices exhibiting six-fold symmetry were observed. The XY piezo constants were calibrated at this position so that the lattice constant matched the theoretical value of 17.7 nm[26,58].

Next, as shown in Fig. 2c, imaging was performed at the upper-right edge of the positioning range (XY offset voltage ≈ 93 V), where the lattice constant was observed to be ~22.3 nm (≈ 25% increase). In contrast, as shown in Fig. 2d, at the lower-left edge (XY offset voltage ≈ 7 V), the lattice constant decreased to ~16.3 nm (≈ 8% decrease). These results indicate that the scan range can differ by up to ~30% between the negative and positive ends. As shown in Fig. 2e, apparent molecular size exhibited a nonlinear dependence and increased steeply at larger offset voltages.

This behavior can be explained as follows. In AFM piezo scanners, ferroelectric domains are pre-aligned by poling, producing macroscopic polarization even at 0 V. Applying a DC offset reduces the contribution of movable domain walls, lowering the local slope of the displacement–voltage curve. As a result, the effective AC piezoelectric coefficient decreases with increasing offset voltage, making the molecular lattice appear artificially enlarged.

To compensate for this effect, the required modulation depth $m$ of the scan-signal amplitude can be approximated by a quadratic function, given by

$$m = 1 + \alpha_{\text{offset}} p_{\text{offset}} + \beta_{\text{offset}} p_{\text{offset}}^2, \quad (1)$$

where the coefficients $\alpha_{\text{offset}}$ and $\beta_{\text{offset}}$ correspond to the linear and quadratic terms, respectively. $p_{\text{offset}}$ represents the normalized XY offset voltage, scaled to the range −1 to +1, which can be expressed by

$$p_{\text{offset}} = 2\frac{V_{\text{offset}}}{V_{\text{Max}}} - 1, \quad (2)$$

where $V_{\text{offset}}$ and $V_{\text{Max}}$ denote the applied offset voltage and the maximum output voltage of the piezo driver (100 V in our setup), respectively. As shown in Fig. 2e, the experimental results are well fitted by Eq. (1).

As these equations indicate, $m = 1$ at the center of the positioning range ($V_{\text{offset}} = 50$ V and $p_{\text{offset}} = 0$). Let $m_-$ and $m_+$ denote the values of $m$ at $p_{\text{offset}} = -1$ and $p_{\text{offset}} = +1$, respectively. The coefficients in Eq. (1) can then be determined as:

$$\alpha_{\text{offset}} = \frac{m_+ - m_-}{2} \quad \text{and} \quad (3)$$

$$\beta_{\text{offset}} = \frac{m_+ + m_-}{2} - 1. \quad (4)$$

Thus, the user first calibrates the XY piezo constants by imaging a calibration sample at the center of the positioning range. Next, imaging is performed near the negative and positive ends of the X positioning range to calibrate $m_-$ and $m_+$, from which $\alpha_{\text{offset}}$ and $\beta_{\text{offset}}$ can be obtained. Since measurement errors are minimized among the three calibration points, it is preferable to perform calibration at positions slightly away from the extreme edges rather than exactly at the edges. The same procedure is applied to the Y direction, enabling correct scan sizes to be set at any position within the XY range. Tests on various similar-type scanners showed that $m_+$ ranges from 1.2 to 1.3 and $m_-$ ranges from 0.85 to 0.95 in our setup.

## 6. Scan size nonlinearity

In AFM measurements, scan sizes are adjusted based on the size of the target structures and the measurement purpose (e.g., whether a single structure or multiple structures are imaged simultaneously). Therefore, in this section, we describe how to ensure that quantitative distance measurements can be performed across a range of scan sizes.

The AC voltage of the scan signal applied to the piezo, $V_{scan}$, is typically associated with the user-defined scan range, $R_{scan}$, assuming a linear piezoelectric constant, as follows:

$$R_{scan} = k_{scan} V_{scan}, \quad (5)$$

where $k_{scan}$ represents the linearly approximated piezoelectric coefficient.

However, actual $R_{scan}$ is not strictly linear with respect to $V_{scan}$. For example, as shown in Fig. 3a, Annexin V was imaged at a scan size of 200 × 200 $nm^2$, and the scanner was calibrated so that the lattice constant matched the ideal value. Under this calibrated condition, imaging at 390 × 390 $nm^2$ resulted in an observed lattice constant of 16.4 nm, approximately 8% smaller (Fig. 3b).

This behavior arises because, even in poled piezo elements, the displacement–electric field characteristics are not strictly linear. As the AC voltage increases, reversible domain-wall motion and rotation of non-180° domains are activated. As a result, the displacement amplitude increases nonlinearly with the electric-field amplitude, and the effective AC piezoelectric coefficient increases with increasing AC voltage.

To evaluate this quantitatively, we measured the AC voltage dependence of the piezo displacement using a laser displacement sensor. As shown in Fig. 3c, the displacement exhibited a nonlinear relationship, with a smaller slope at low voltages and a steeper slope at high voltages. Figure 3d shows the deviation from a linear approximation assuming calibration was performed at 200 nm. As seen, small scan sizes exhibited errors of up to ~7 %, whereas large scan sizes can show deviations of up to ~25 %.

Based on these results, we adopted the following quadratic expression to estimate the actual scan size:

$$R_{scan} = \alpha_{scan}\tilde{V}_{scan}\left(1+\beta_{scan}\tilde{V}_{scan}\right), \tag{6}$$

where $\tilde{V}_{scan}$ is the normalized $V_{scan}$, defined as follows:

$$\tilde{V}_{scan} = \frac{V_{scan}}{V_{Max}}. \tag{7}$$

$\alpha_{scan}$ and $\beta_{scan}$ represent the scaling factor and nonlinear factor of the piezoelectric coefficient, respectively, where $\beta_{scan} = 0$ corresponds to a perfectly linear response. As shown in Fig. 3e, fitting the experimental results using Eq. (6) yielded a good agreement across the entire scan size with $\beta_{scan} = 0.53$.

Testing several scanners revealed that their typical values range from 0.5 to 1 in our setup. Therefore, for example, when using a scanner calibrated at 200 nm to perform a scan at twice that size (400 nm), without any correction, the actual scan size would be 436 nm for $\beta_{scan} = 0.5$ and 467 nm for $\beta_{scan} = 1$. These correspond to scaling errors of 9% and 17%, respectively.

To perform calibration based on this quadratic function, users only need to set $k_{scan}$ and $\beta_{scan}$ in the software during calibration, while $\alpha_{scan}$ is automatically determined as an internal parameter as follows:

$$\alpha_{scan} = \frac{k_{scan}}{1+\beta_{scan}\dfrac{R_{base}}{k_{scan}}}, \tag{8}$$

where $R_{base}$ is a base scan size used as a reference for calibration. Because the positional accuracy becomes highest near the chosen $R_{base}$, it is preferable to select a scan size that is most frequently used, e.g., $200 \times 200\ \text{nm}^2$ for biomolecular imaging.

First, the calibration sample is imaged at $R_{base}$, using the frame rate and pixel numbers typically used for practical imaging. At this stage, the initial value of $\beta_{scan}$ is set to a typical value. Through this measurement, the optimal $k_{scan}$ is calibrated. The determined $k_{scan}$ value is compatible with conventional systems that assume a linear piezoelectric coefficient.

Subsequently, the scan size is changed to a value different from $R_{base}$, for example 600 × 600 $nm^2$, and $\beta_{scan}$ is further optimized. The frame rate is adjusted to maintain the same scan velocity as at $R_{base}$. This calibration procedure enables the determination of both $\alpha_{scan}$ and $\beta_{scan}$. Although the scan accuracy is maximized at the two scan sizes used for calibration, extrapolation of the quadratic function of Eq. (6) to other scan ranges allows high scan accuracy to be maintained over a broader range of scan sizes than with a linear approximation.

To implement this method in software, $V_{scan}$ must be computed inversely from the user-specified $R_{scan}$. By solving Eq. (6) for $\tilde{V}_{scan}$, an analytical expression for $\tilde{V}_{scan}$ as a function of $R_{scan}$ is obtained as follows:

$$\tilde{V}_{scan} = \frac{1}{2\beta_{scan}}\left(\sqrt{1+\frac{4\beta_{scan}R_{scan}}{\alpha_{scan}}}-1\right). \tag{9}$$

Introducing this quadratic correction alone already allows for higher positional accuracy compared with the conventional linear approximation (Fig. 3e). However, as shown in the same figure, the experimental result showed that at large scan sizes, the slope becomes slightly shallower than that predicted by the quadratic function. Consequently, approximating the entire scan range with a quadratic function can result in deviations of up to 6% at large scan ranges. One possible solution is to use a cubic function; however, its inverse, although conditionally obtainable, does not admit a simple closed form like a quadratic function.

Therefore, we adopted a hybrid approach: for $\tilde{V}_{scan}$ below a set normalized threshold $\tilde{V}_{th}$, the quadratic functions given in Eqs. (6) and (9) are used, while for $\tilde{V}_{scan}$ above $\tilde{V}_{th}$, a linear extrapolation of the quadratic function is employed as follows:

$$R_{scan} = \alpha_{scan}\left[\left(1+2\beta_{scan}\tilde{V}_{th}\right)\tilde{V}_{scan} - \beta_{scan}\tilde{V}_{th}^{2}\right] \quad \text{for } \tilde{V}_{scan} > \tilde{V}_{th}. \tag{10}$$

The scan size $R_{th}$ corresponding to $\tilde{V}_{th}$ is given by:

$$R_{th} = \alpha_{scan}\tilde{V}_{th}\left(1+\beta_{scan}\tilde{V}_{th}\right). \tag{11}$$

Furthermore, the inverse of Eq. (10) can be solved analytically as shown below, facilitating straightforward software implementation:

$$\tilde{V}_{\text{scan}} = \frac{R_{\text{scan}} + \alpha_{\text{scan}}\beta_{\text{scan}}\tilde{V}_{\text{th}}^{2}}{\alpha_{\text{scan}}\left(1 + 2\beta_{\text{scan}}\tilde{V}_{\text{th}}\right)} \quad \text{for } \tilde{V}_{\text{scan}} > \tilde{V}_{\text{th}}. \tag{12}$$

In our case, the typical threshold voltage was found to be $\tilde{V}_{\text{th}} \approx 0.5$. With this correction, as shown in Fig. 3f, higher-accuracy positional control can be maintained across the entire scan range.

## 7. Scan wave hysteresis – principle

In piezo scanners, the displacement–voltage relationship is relatively linear at small AC voltages, whereas at large AC voltages it becomes nonlinear due to significant domain-wall motion (Fig. 1b). Because this motion lags behind the applied voltage, the nonlinearity is accompanied by hysteresis[5,31,56,59]. In this section, we describe an analytical compensation model for this hysteretic behavior.

Figure 4 summarizes the concept of piezo hysteresis compensation. In typical AFM imaging, a linear waveform is used as the scan signal, which can be expressed as follows:

$$V(t) = V_{\text{scan}} \tau_i(t), \tag{13}$$

where $\tau$ denotes the normalized ideal linear waveform, expressed as follows:

$$\tau_i = \begin{cases} \dfrac{t}{\alpha_{\text{turn}}} & \text{for } 0 < t \le \alpha_{\text{turn}} \left(i = 1\right), \\ 1 - \dfrac{t - \alpha_{\text{turn}}}{1 - \alpha_{\text{turn}}} & \text{for } \alpha_{\text{turn}} < t \le 1 \left(i = 2\right), \end{cases} \tag{14}$$

varying from 0 to 1 in the forward scan and from 1 to 0 in the backward scan. Here, $t$ represents the normalized time of a single line-scan period, and $\alpha_{\text{turn}}$ denotes its turning point between forward and backward scans. An $\alpha_{\text{turn}}$ value of 0.5 corresponds to a standard triangular-wave X scan, whereas values greater than 0.5 produce a sawtooth-like waveform.

When a large-amplitude signal obtained from Eq. (14) is applied to the piezo, the resulting displacement response is distorted due to the piezo's nonlinear transfer function, denoted $G_{\text{piezo}}$ (Fig. 4a). Scanning with this distorted waveform thus produces position-dependent scaling errors in the AFM image: during the forward scan, the left half appears stretched while the right half is compressed, whereas the opposite occurs in the backward scan (Fig. 4b). Consequently, a feature near the image center appears shifted to the left in the backward scan relative to the forward scan (see the purple dashed circle and reference line). The transition between stretching and compression occurs near the center in the backward scan but is slightly left-shifted in the forward scan. Similar nonlinear behavior is also observed in the Y-scan direction.

An ideal linear piezo response can be achieved by applying the inverse piezo transfer characteristic to the desired linear waveform as described in detail in the next section (Fig. 4c). Imaging with this corrected waveform yields AFM images with uniform scaling across the entire scan area (Fig. 4d).

Additionally, suppression of high-frequency components is essential to prevent excitation of the piezo resonances[10,13], which can result in vertical streaks (ringing) in AFM images. To generate a nonlinear waveform while achieving this suppression, it is important to construct the waveform from its Fourier series representation. This approach provides a low-pass filter (LPF) with an ideal frequency response and a flexible cutoff frequency, and can also be readily extended to inverse-compensation-based damping[12,13,47].

Prior to discussing nonlinear waveform generation, for reference, we experimentally measured the displacement–voltage curve obtained by applying a 100 V triangular voltage to the scanner [Fig. 5a]. As shown in the upper panel, the forward and backward responses follow different paths, resulting in a characteristic hysteresis. To quantitatively evaluate the deviation from the ideal linear response (gray dashed line), the residuals are plotted in the lower panel. The residuals peak at 3.5% and 8.5% for the forward and backward scans, respectively, indicating that the nonlinearity in the backward scan is more than two times larger than that in the forward scan. The same characteristics were also observed in piezo scanners with an open-ended structure, suggesting that it is not intrinsic to flexure structures but rather to stacked piezo actuators.

In addition, the peak positions differ slightly between the forward and backward scans. The peak is located at approximately 0.35 in the forward scan, whereas it appears at around 0.55 in the backward scan. Because the backward scan is performed in the opposite sweep direction to the forward scan, the peak position corresponds to approximately 0.45 when the sweep directions are aligned. Consequently, in both cases, the asymmetry-related peak is found at a value slightly smaller than 0.5.

## 8. Scan wave hysteresis – compensation

In this section, we consider waveforms for compensating piezo hysteresis using an analytical approximation method. The nonlinearity of the inverse piezo response can be approximated by a linear waveform with a superimposed sine waveform[52], expressed as:

$$u(\tau_i) = \begin{cases} \tau_1 + \beta_1 \sin\left(\pi \tau_1\right) & \text{for } 0 < t \leq \alpha_{\text{turn}}, \\ \tau_2 - \beta_2 \sin\left(\pi \tau_2\right) & \text{for } \alpha_{\text{turn}} < t \leq 1, \end{cases} \tag{15}$$

where $u$ denotes the normalized nonlinear waveform, varying from 0 to 1 in the forward scan and from 1 to 0 in the backward scan. $\beta_1$ and $\beta_2$ represent the nonlinear factors of the forward and backward scan waveforms, respectively. Hereafter, the parameter pair ($\beta_1$, $\beta_2$) is collectively denoted as $\beta_i$ with $i$ = 1, 2. This expression can be Fourier-transformed (see Supplementary Note 1), making it applicable to practical implementation. Hereafter, we refer to this expression as the sinusoidal model and first analyze compensation using this approach.

To align the axes with the experimental results (Fig. 5a), the inverse transfer characteristic of Eq. (16) is plotted in Fig. 5b. As shown, the nonlinearity increases with $\beta_i$. The peak residuals reach −3%, −6%, and −9% for $\beta_i$ = 0.03, 0.06, and 0.9, respectively, indicating that the magnitude of $\beta_i$ directly corresponds to the peak residual. However, the peak positions of the residuals increase from 0.53 to 0.56 and 0.59 with increasing $\beta_i$, indicating that they shift toward the right side of the center. Analysis of the inverse transfer characteristic in Eq. (15) (Supplementary Note 2) shows that the peak position $u_{\text{peak},i}$ is given by

$$u_{\text{peak},i} = \frac{1}{2} + \beta_i. \tag{16}$$

Consequently, the peak position in the sinusoidal model is always greater than or equal to 0.5. This asymmetry is opposite to the experimental observations, in which the peak appears on the left side of the center (i.e., less than 0.5).

To evaluate the extent to which the piezo nonlinearity can be compensated, we analyzed the time-domain waveforms of the input signal and the piezo response. The displacement of the piezo scanner was characterized using a laser displacement sensor. As a reference, applying an uncorrected triangular wave resulted in large deviations from the ideal linear waveform, with peaks of −3.5% and +8.5% for forward and backward scans, respectively (Fig. 5c). In contrast, applying a waveform generated by the sinusoidal model with an optimized $\beta_i$ yielded a waveform closely matching a triangular wave (upper panel of Fig. 5d). However, as shown in the lower panel, the residual did not completely vanish and remained at approximately ±1.5% at the peak, because the asymmetry of the nonlinearity differed from that observed experimentally.

Next, we analyze the extent to which these errors affect the image scaling accuracy. Since the effective piezoelectric coefficient corresponds to the slope of the piezo response, it can be derived by differentiating the residual curves. As an approximation, a maximum scaling error is about six times the peak amplitude of the residual curve (Supplementary Note 3).

Figure 5e shows the scaling error obtained by differentiating the experimentally measured residual curve, plotted as a function of the normalized $x$-position in the scan. A positive error indicates that the image appears stretched, whereas a negative error indicates compression. Calibration is typically performed so that the scaling at the center of the image is correct; therefore, the offset is adjusted such that the error at $x = 0.5$ is displayed as 0%.

In practice, quantitative image analysis is often limited to structures near the image center, because image distortion tends to increase toward the edges due to LPF effects. Focusing on the central region only ($x = 0.1$–$0.9$), i.e., the region not shaded in green, we find that when a triangular waveform is used, the maximum peak-to-peak scaling error in the forward scan reaches approximately 30%. Moreover, the error is pronounced mainly in the left half of the image, while the right half shows relatively little deviation. In contrast, in the backward scan the error increases markedly toward the edges, reaching a maximum peak-to-peak amplitude of about 60%.

By comparison, when the sinusoidal model is used, the maximum peak-to-peak scaling error is reduced to approximately 15% for both forward and backward scans (Fig. 5f). In addition, the error profile is relatively flat near the image center, indicating that scaling errors are more effectively suppressed in this region.

To better align the model's nonlinearity with the actual piezo response, we derived a model that accounts for asymmetry by adding a cosine term to the sinusoidal model. This modification, which superimposes only first-harmonic components, incorporates nonlinearity without introducing higher-frequency components. We refer to this as the harmonic model, expressed as follows (see Supplementary Note 4 for the derivation):

$$u(t)=\begin{cases}\lambda_1\tau_1+\xi_1\left[\sin\left(\pi\tau_1\right)+\gamma_1\left(\cos\left(\pi\tau_1\right)-1\right)\right] & \text{for } 0<t\le\alpha_{\text{turn}},\\ \lambda_2\tau_2+\xi_2\left[-\sin\left(\pi\tau_2\right)+\gamma_2\left(\cos\left(\pi\tau_2\right)-1\right)\right] & \text{for } \alpha_{\text{turn}}<t\le 1.\end{cases} \tag{17}$$

By considering the boundary conditions for $\tau$ and $u$, $\lambda_i$ is obtained as follows:

$$\lambda_i=1+2\xi_i\gamma_i. \tag{18}$$

Furthermore, the expressions for the nonlinearity parameter $\xi_i$, and the asymmetry parameter $\gamma_i$ in terms of $\beta_i$ and $u_{\text{peak},i}$ are given as follows:

$$\xi_i=\frac{\beta_i}{\gamma_i\left[\cos\left(\pi u_{\text{peak},i}^{(0)}\right)+2u_{\text{peak},i}^{(0)}-1\right]+\sin\left(\pi u_{\text{peak},i}^{(0)}\right)}, \tag{19}$$

$$\gamma_i=\frac{\pi}{2}\,\frac{\tan^2\left(\frac{\pi u_{\text{peak},i}^{(0)}}{2}\right)-1}{\tan^2\left(\frac{\pi u_{\text{peak},i}^{(0)}}{2}\right)-\pi\tan\left(\frac{\pi u_{\text{peak},i}^{(0)}}{2}\right)+1}, \tag{20}$$

where $u_{\text{peak},i}^{(0)}$ represents $u_{\text{peak},i}$ in the limit as $\beta_i \rightarrow 0$ and is expressed as follows:

$$u_{\text{peak},i}^{(0)}=u_{\text{peak},i}-\beta_i. \tag{21}$$

As shown in Fig. 5a, the peak residuals for the forward and backward scans have inherently opposite signs, and $u_{\text{peak}}$ is likewise mirrored, taking the value $(1-u)$. However, in Eqs. (17)–(21), $\beta$ and $u$ in the backward scan is expressed in the same coordinate system as the forward scan, so these inversions are unnecessary. For practical implementation, Eq. (17) can be expanded as a Fourier series (Supplementary Note 1).

To investigate the characteristics of this model, we first analyze waveforms for different values of $\beta_i$ with $u_{\mathrm{peak},i}$ = 0.48 (Fig. 5g). As shown, the nonlinearity increases with $\beta_i$, while the residual curves show that the peak position remains at 0.48, while the peak residual increases proportionally with $\beta_i$. For $\beta_i$ = 0.03, 0.06, and 0.09, the peak values are exactly −3%, −6%, and −9%, respectively, indicating that, as in the sinusoidal model, the value of $\beta_i$ directly corresponds to the peak amplitude.

Next, we examine waveforms for different values of $u_{\mathrm{peak},i}$ with $\beta_i$ = 0.035 (Fig. 5h). As shown in the upper panel, all curves nearly overlap. In contrast, although all residual curves have a peak residual of −3.5%, their peak positions differ and coincide exactly with the specified values ($u_{\mathrm{peak},i}$ = 0.37, 0.5, and 0.65). These results demonstrate that Eqs. (17)–(21) accurately reproduce the prescribed parameters. Furthermore, this indicates that, unlike the sinusoidal model, $u_{\mathrm{peak},i}$ can be reduced to less than 1/2 of the value corresponding to the actual piezo response.

Experimentally, we investigated the dependence of the scan voltage on $u_{\mathrm{peak},i}$ and found that it does not vary significantly with voltage. Likewise, piezo elements of the same model exhibited little device-to-device variation. Even if $u_{\mathrm{peak},i}$ varies slightly with the element design, optimizing $u_{\mathrm{peak},i}$ allows the hysteresis compensation to be tuned for each device. Therefore, compared with the sinusoidal model, the harmonic model offers greater versatility.

To evaluate piezo nonlinearity compensation, we measured the time-domain response of the piezo upon application of waveforms generated using the harmonic model (Fig. 5i). The optimal parameter sets ($\beta_i$, $u_{\mathrm{peak},i}$) were determined to be (0.035, 0.37) for the forward scan and (0.085, 0.48) for the backward scan. In the upper panel, compared with the sinusoidal model, the forward waveform showed a slightly steeper rise around 5 s. In the lower panel, the residuals revealed that the peak value in the forward direction is substantially reduced to about one third of that of the sinusoidal model (approximately ±0.5%). In contrast, although the backward residual was also reduced but did not completely vanish due to a slight mismatch in peak shape (approximately ±1% at the peak).

Analysis of the scaling error indicates that the peak-to-peak amplitude was about 11% in the backward scan, while it was reduced to roughly 5% in the forward scan (Fig. 5j). This demonstrates that the harmonic model can significantly suppress nonlinearity, particularly in the forward scan. A topographic image acquired using the corrected scan waveform is shown in Fig. 4(d), where the grid structure was observed to be uniformly spaced across the entire image.

## 9. Scan wave hysteresis – size dependence

Since piezo hysteresis is voltage-dependent, in this section, we examine how the piezo nonlinearity varies with scan size. Triangular waveforms corresponding to different voltages (scan sizes) were applied to the piezo, with the scan frequency adjusted so that the scan velocity matched the typical HS-AFM value of 100 μm/s. As the hysteresis increases slightly with increasing frequency[43,60,61], it is important to use the typical scan frequencies employed in actual experiments for each scan size, rather than a fixed frequency. The absolute values of the peak residuals for the forward and backward scans are plotted in Fig. 5k against $\tilde{V}_{\text{scan}}$.

As shown, the peak residuals for both forward and backward scans increase with $\tilde{V}_{\text{scan}}$; however, at larger scan sizes, this trend saturates, resulting in a more gradual increase. Comparing the two curves, no significant difference is observed at small scan sizes, but as the scan size grows, the nonlinearity in the backward scan becomes more pronounced. For example, without compensation, at a scan size of 200 nm, the distance error in AFM images for the forward scan is approximately 2–3%, whereas it can approach 10% at larger scan sizes.

Furthermore, we found that this behavior could be well described by the exponential function given as follows:

$$\beta_i = \beta_i^{\max} \frac{1-\exp\left(-\rho_{\text{scan}}\tilde{V}_{\text{scan}}\right)}{1-\exp\left(-\rho_{\text{scan}}\right)}, \tag{22}$$

where $\beta_i^{\max}$ and $\rho_{\text{scan}}$ represent the maximum $\beta_i$ and decay constant, respectively. When $\tilde{V}_{\text{scan}} = 0$, $\beta_i = 0$, and when $\tilde{V}_{\text{scan}} = 1$, $\beta_i = \beta_i^{\max}$. For example, in Fig. 5k, the forward and backward results were fitted with the dashed lines using parameters of $\beta_i^{\max} = 0.037$, $\rho_{\text{scan}} = 4\ \text{nm}^{-1}$ and $\beta_i^{\max} = 0.08$, $\rho_{\text{scan}} = 1.5\ \text{nm}^{-1}$, respectively. Therefore, the nonlinear factor at an arbitrary scan size can be determined by fitting Eq. (22) to nonlinear factors measured at only a few representative scan sizes.

Although triangular waveforms ($\alpha_{\text{turn}} = 0.5$) are typically used for standard scanning, sawtooth waveforms ($\alpha_{turn}$ > 0.5) are sometimes employed to increase the scan speed. In this case, images are acquired only during the forward scan, while the backward scan is skipped by temporarily lifting the probe from the sample surface[55,62,63]. Accordingly, we performed a similar analysis for different values of $\alpha_{turn}$. As shown in Fig. 5l, varying $\alpha_{turn}$ does not generally produce major differences. However, at larger scan sizes, the nonlinearity increased to approximately 10% for $\alpha_{turn}$ = 0.7 and 25% for $\alpha_{turn}$ = 0.9.

Furthermore, the Y piezo also exhibits nonlinearity, similar to the X piezo. For Y scanning, images can be acquired in both the upward and downward directions ($\alpha_{turn}$ = 0.5). However, in this scanning mode, drift-induced image distortion reverses from frame to frame because the drift direction is opposite between upward and downward scans. Therefore, for dynamic observations, especially HS-AFM imaging, it is preferable to record images in only one direction using a sawtooth-like waveform ($\alpha_{turn} \lesssim 1$). Accordingly, a similar analysis was performed for the Y scanner in this configuration, revealing that the characteristics of $\beta_1$ and $u_{peak,1}$ are consistent with those observed in the forward scan of the X scanner. Although sawtooth-like scanning tends to increase nonlinearity, the hysteresis does not become markedly larger in Y scanning because the scanning frequency is lower. These results indicate that the proposed correction model is applicable to both X and Y scanners.

Finally, we summarize practical approaches to minimize distance errors arising from hysteresis. Regardless of whether compensation is applied, the scaling error in forward scans is significantly smaller than that in backward scans. Furthermore, since sufficient compensation cannot always be guaranteed under all measurement conditions, forward scans generally provide more stable suppression of scaling errors. Since hysteresis increases with scan size, it can be minimized by restricting the scan range to values well below the maximum. Moreover, distance errors are smallest near the image center; therefore, restricting the analysis to central structures is important for further error reduction.

## 10. Scan frequency

In AFM imaging, even for the same scan size, the frame rate must be adjusted according to the dynamical behavior and fragility of the sample. However, the apparent AC piezoelectric coefficient exhibits a slight frequency dependence. For example, as shown in Fig. 6a, Annexin V was imaged at a scan size of 100 × 100 nm$^2$ and 0.5 fps, and the scanner was calibrated under this condition so that the lattice constant became 17.7 nm. When the frame rate was subsequently increased by a factor of 14 to 7.3 fps, the apparent lattice constant increased to 18.3 nm, corresponding to a 3% increase (Fig. 6b).

This phenomenon originates from rate-dependent hysteresis, in which piezo hysteresis becomes more pronounced and the maximum displacement slightly decreases with increasing frequency due to the dynamic loss of domain-wall motion[43,60,61]. Consequently, the apparent molecular size increases with increasing frame rate, even for the same scan size.

To evaluate this effect quantitatively, the frequency gain response of the X piezo was measured. As shown in Fig. 6c, the gain was nearly flat at low frequencies, but a resonance peak appeared at approximately 50 kHz. However, when the vertical axis is expanded (Fig. 6d), a slight decrease in gain with frequency was observed even in the low-frequency range. This frequency characteristics was well fitted using the following empirical formula:

$$R_{\text{scan}}(f_{\text{scan}}) \approx R_{\text{scan}}(f_{\text{DH}}) - \alpha_{\text{DL}} \log_{10}\left(\frac{f_{\text{scan}}}{f_{\text{DH}}}\right), \tag{23}$$

where $f_{\text{scan}}$, $f_{\text{DL}}$, and $\alpha_{\text{DL}}$ denote the scan frequency, the reference frequency, and the logarithmic slope associated with dynamic hysteresis, respectively. A typical value of $\alpha_{\text{DL}}$ was found to be 0.03, corresponding to an approximately 3% decrease per decade.

Calibration of scan hysteresis and scan-frequency dependence requires a laser displacement sensor, which is generally unavailable in laboratories that do not develop AFM instruments. However, because the scan speed is typically adjusted only within a factor of two to three, frequency-dependent errors remain below 1−2%. Furthermore, this frequency dependence shows no significant voltage dependence, and tests on multiple piezo scanners reveal no significant differences, suggesting that typical values can be used as common parameters.

## 11. Nonlinearity in Z scanner

Similar to the XY scanner, the Z scanner can exhibit piezo nonlinearity, which can lead to errors in height analysis. However, because the Z scanner operates under feedback control during AFM imaging, the feedforward compensation method cannot be directly applied. Therefore, in this section, we discuss the extent to which Z-scanner nonlinearity affects quantitative measurements.

Unlike XY scanners, the Z scanner has an open-ended structure and is subject to minimal mechanical loading. Therefore, we compared the nonlinearities of piezo elements in flexure-guided and open-ended structures but found no significant differences, suggesting that the Z scanner exhibits positional errors comparable to those of XY scanners. However, the positioning errors of Z piezo are expected to be relatively small compared to the XY scanner for the following reasons.

First, to avoid image loss or tip–sample crashes, the Z offset voltage in typical AFM measurements is usually limited to near the center of the travel range. As a result, the offset-voltage dependence, which strongly contributes to piezo nonlinearity, is much smaller for the Z scanner. For the same reason, during practical imaging, the Z scanner is operated over a limited displacement range, typically well below half of its maximum travel. This operating condition effectively suppresses hysteresis and the voltage dependence of the AC piezoelectric coefficient.

Furthermore, when observing samples with small height variations such as biomolecules, the required Z range is limited, allowing the use of a low-voltage piezo driver and thereby suppressing piezo nonlinearity. For example, in our HS-AFM system, a 50 V piezo driver is typically used, reducing the nonlinearity to less than one third compared with a 100 V range. Overall, even in open-loop operation, the influence of nonlinearity in the Z scanner is smaller than that in the XY piezo.

## 12. Conclusions

In this study, we identified four types of nonlinear effects in open-loop piezo AFM scanners and experimentally assessed their impact on positioning errors. We developed simple analytical models to compensate for these effects, achieving high-precision control without specialized hardware. The software-based correction preserves scanner bandwidth and noise characteristics, ensuring compatibility with existing systems. For high-resolution imaging, tip-position (offset voltage) dependence and scan-size nonlinearity are the most significant factors, both of which can be readily calibrated using standard imaging-based procedures. This approach is applicable to both stack-type and tube piezos, making it also compatible with other scanning probe microscopies. Although it applies only to the XY scanner and not to the Z scanner under feedback control, scaling errors in the Z scanner are generally smaller, rendering this a minor concern. The method has already been successfully employed in recent practical biomolecular studies using HS-AFM[15-22,24,29,30], demonstrating its effectiveness for extracting quantitative structural information and highlighting its growing importance for future high-precision dynamics imaging applications.

## Acknowledgements

We thank Professor Toshio Ando of Kanazawa University for valuable support, and the technical assistants Ms. Aimi Makino and Ms. Kayo Nakatani for calibrating the piezo scanners. This work was supported by PRESTO, Japan Science and Technology Agency (JST) [JPMJPR20E3 and JPMJPR23J2 to K.U.]; and KAKENHI, Japan Society for the Promotion of Science [25K09575 (to K.U.), and 24H00402 (to N.K.)].

## Author contributions

K. U. constructed the theories, performing the experiments, and wrote the manuscript; and N. K. supervised the study.

## Conflict of Interest

The authors serve as technical advisors to Research Institute of Biomolecule Metrology Co. Ltd., and the functionality described in this paper has been incorporated into commercial instruments manufactured by the company.

## Data availability statement

The datasets generated and/or analyzed during the current study are available from the corresponding author on reasonable request.

## Figures

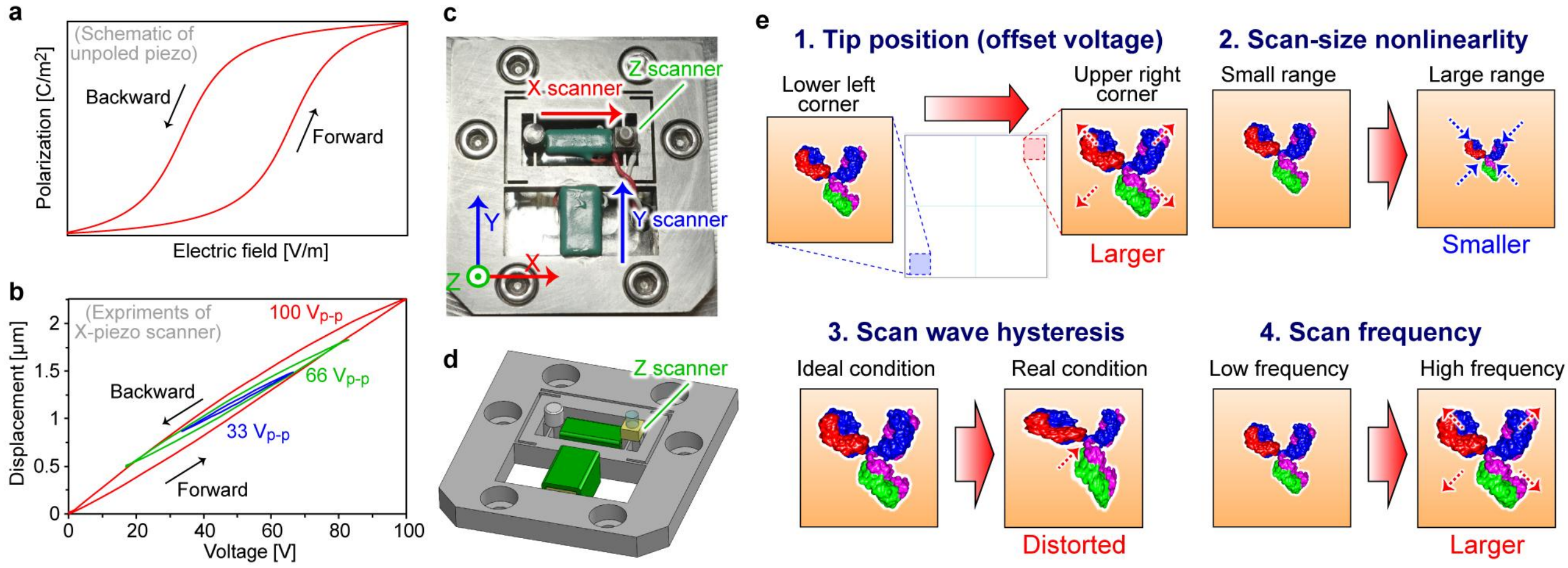

**Fig. 1.** (**a**) Schematic of polarization–electric field characteristics for an unpoled piezo element, which is approximated by a hyperbolic tangent function. (**b**) Experimental displacement–voltage characteristics for the AFM X-piezo scanner shown in panel (c), measured using triangular voltage waveforms over voltage ranges of 33, 66, and 100 V. (**c,d**) Photograph (c) and 3D CAD model (d) of the HS-AFM scanner used in the experiments. (**e**) Summary schematics illustrating four types of nonlinearities in piezo scanners that affect AFM imaging. A monoclonal antibody (PDB: 1IGT) is shown as a schematic model, with its apparent size and shape exaggerated for clarity.

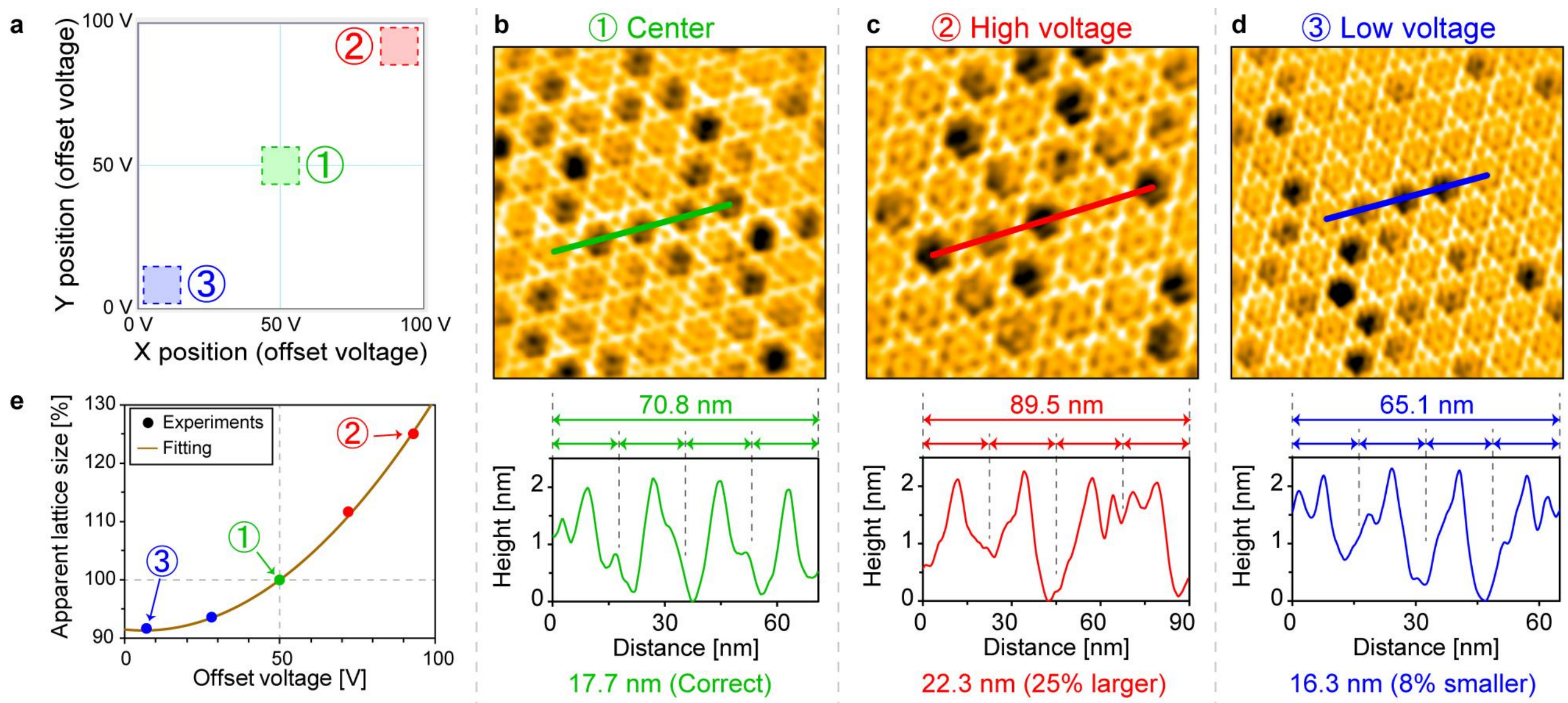

**Fig. 2.** (**a**) Representative AFM control software interface for controlling the XY tip position (offset voltage). (**b–d**) Representative AFM images (upper) and corresponding line profiles (lower) of an Annexin V 2D crystal acquired at different tip positions, as indicated in panel (a). AFM images were acquired with a nominal scan size of 200 × 200 nm$^2$ at 1.8 fps and are shown as magnified views of 130 × 130 nm$^2$. (**e**) Representative experimental dependence of the apparent size scaling on the offset voltage; this scaling is equivalent to the modulation depth required to compensate the scaling error. The orange curve shows the fit obtained using Eq. (1).

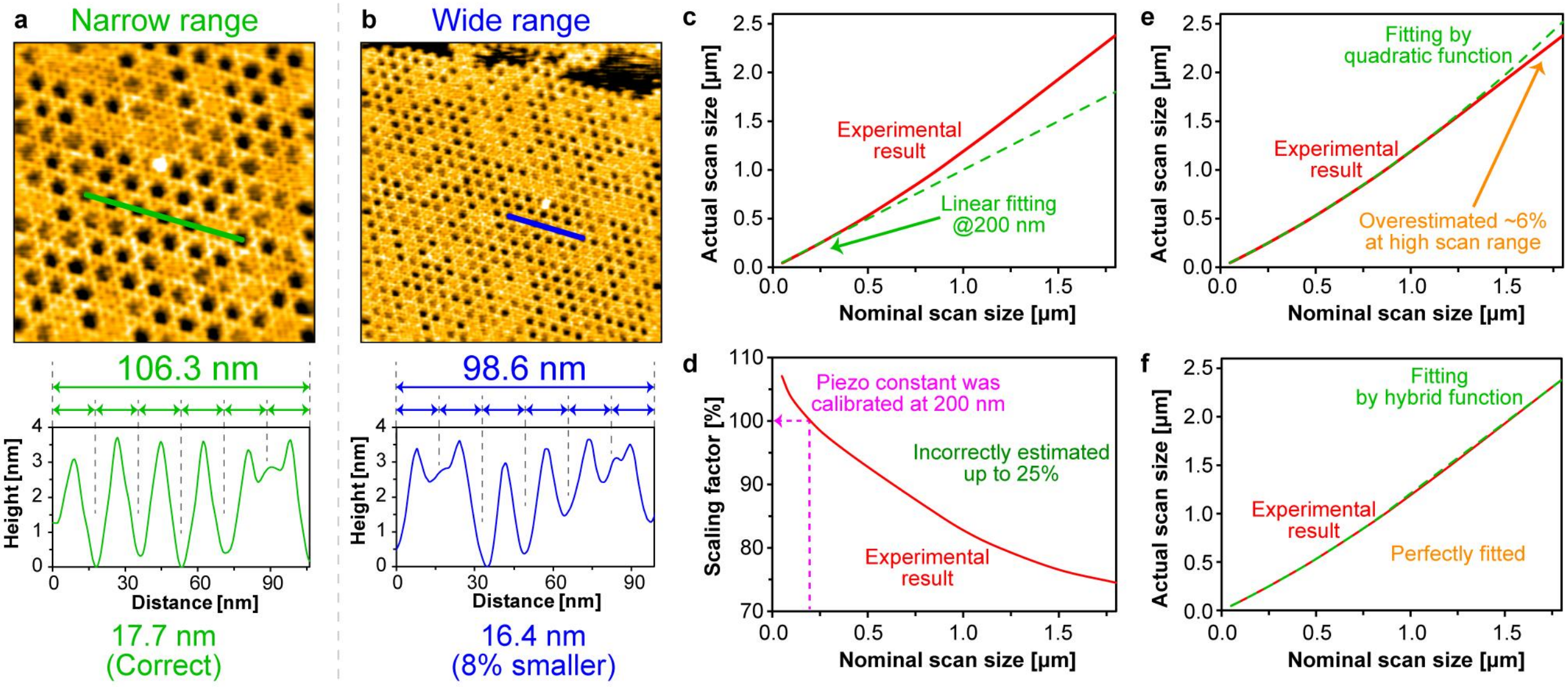

**Fig. 3.** (**a,b**) Scan-size dependence of representative AFM images (upper) and corresponding line profiles (lower) of the Annexin V 2D crystal with a small (a, 200 × 200 nm², 1.8 fps) and a large (b, 390 × 390 nm², 0.7 fps) scan sizes. (**c**) Correlation between the nominal scan size, set under the assumption of a linear piezoelectric coefficient, and the actual scan size determined by the laser displacement sensor. (**d**) Scan-size dependence of the scaling factor required to obtain the correct scan size. (**e,f**) Scan sizes predicted by the quadratic model (e) and the hybrid linear–quadratic model (f), compared with experimental results.

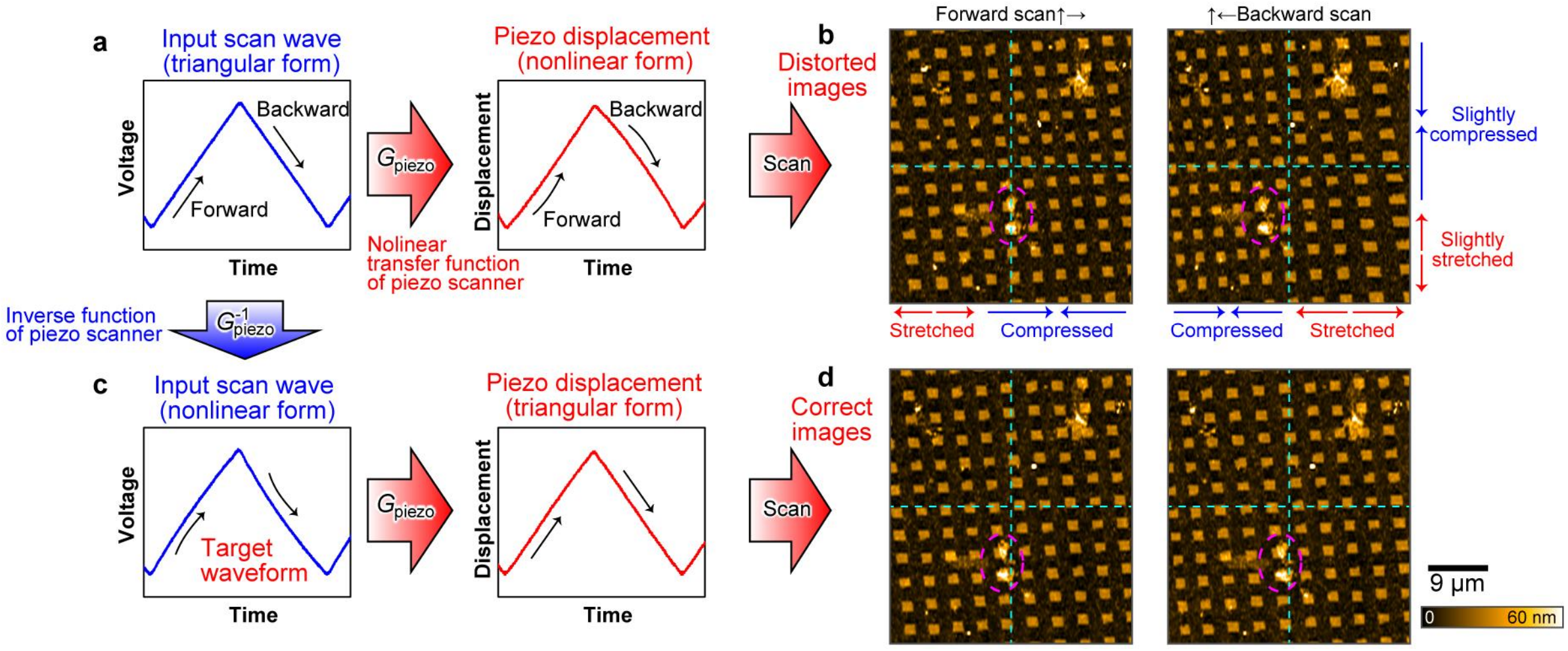

**Fig. 4.** (**a,c**) Schematics of the input scan waveforms and the resulting piezo displacements obtained using a standard triangular waveform (a) and an inverse-corrected waveform (c). (**b,d**) Representative AFM images of a grating sample obtained using a standard triangular waveform (b) and an inverse-corrected waveform (d). AFM images were acquired with a maximum scan size of 31.3 × 35.3 μm$^2$ at 0.008 fps. In each panel, the left and right images correspond to the forward scan (left-to-right direction) and backward scan (right-to-left direction), respectively. The Y-direction scan was performed using a sawtooth-like waveform scanning from bottom to top.

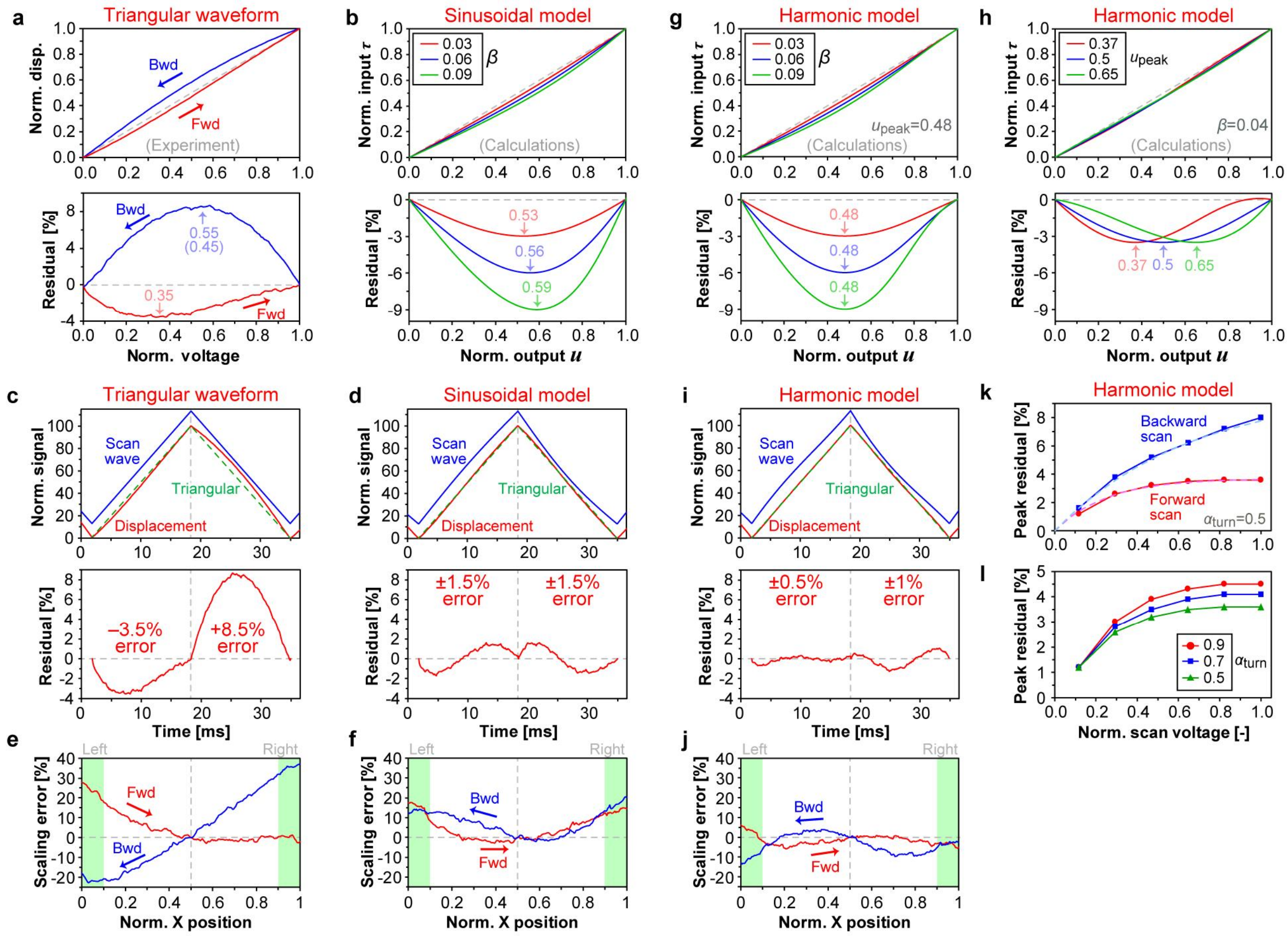

**Fig. 5.** (**a**) Comparison of the forward and backward piezo responses obtained experimentally with a triangular voltage input (upper), together with the corresponding residual curve (lower). (**b,g,h**) Dependence of the inverse characteristics of nonlinear functions on the nonlinear factor $\beta_i$ for the sinusoidal model (b) and on $\beta_i$ (g) and the asymmetric factor $\gamma_i$ (h) for the harmonic model (upper), together with the corresponding residual curves (lower). (**c,d,i**) Time trajectories of scan waveforms and the piezo displacement responses obtained using triangular (c), sinusoidal-model (d), and harmonic-model (i) input waveforms (upper), together with their residuals from the ideal triangular waveform (lower). (**e,f,j**) Scaling error in scanning as a function of normalized x position obtained using triangular (e), sinusoidal-model (f), and harmonic-model (j) input waveforms. (**k,l**) Scan-size dependence of the peak residual obtained with a triangular waveform input, comparing forward and backward scans (k) and different $\alpha_{turn}$ values (l), with dashed lines indicating theoretical fits.

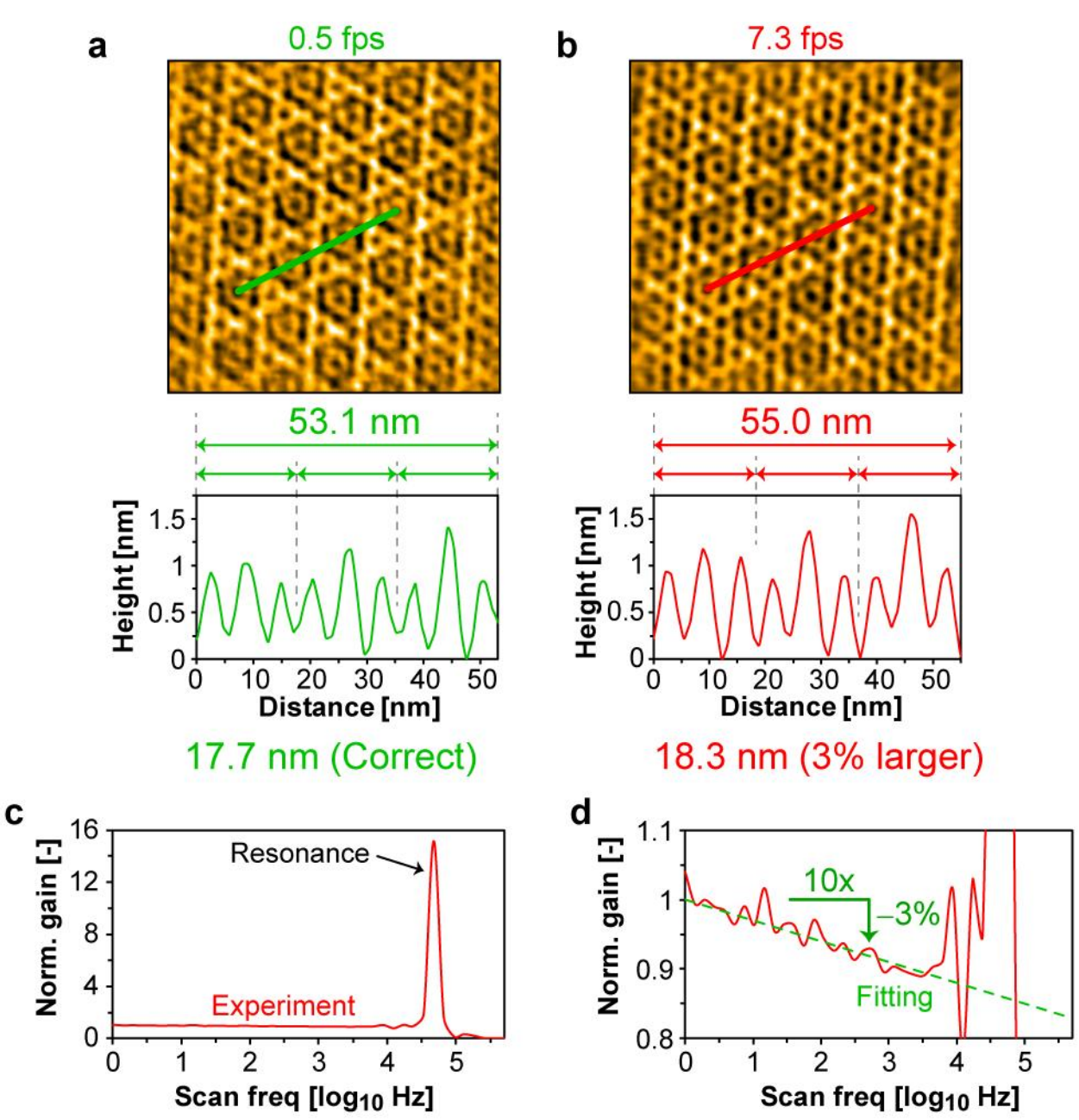

**Fig. 6.** (**a,b**) Scan-frequency dependence of representative AFM images (upper) and corresponding line profiles (lower) of the Annexin V 2D crystal at lower (a, 0.5 fps) and higher (b, 7.3 fps) imaging rates. Both of the images were acquired at a nominal scan size of 100 × 100 $nm^2$. (**c,d**) Frequency dependence of the piezo AC response measured at an AC voltage of 1.6 $V_{p\text{–}p}$ (c) and a magnified view of the data (d).

## References

1. Binnig, G., Quate, C. F. & Gerber, C. Atomic Force Microscope. *Phys. Rev. Lett.* **56**, 930-933 (1986).
2. Garcia, R. & Perez, R. Dynamic atomic force microscopy methods. *Surf. Sci. Rep.* **47**, 197-301 (2002).
3. Giessibl, F. J. Advances in atomic force microscopy. *Rev. Mod. Phys.* **75**, 949-983 (2003).
4. Dufrêne, Y. F., Ando, T., Garcia, R., Alsteens, D., Martinez-Martin, D., Engel, A., Gerber, C. & Müller, D. J. Imaging modes of atomic force microscopy for application in molecular and cell biology. *Nat. Nanotechnol.* **12**, 295-307 (2017).
5. Habibullah, H. 30 Years of atomic force microscopy: Creep, hysteresis, cross-coupling, and vibration problems of piezoelectric tube scanners. *Measurement* **159**, 107776 (2020).
6. Korolkov, V. V., Summerfield, A., Murphy, A., Amabilino, D. B., Watanabe, K., Taniguchi, T. & Beton, P. H. Ultra-high resolution imaging of thin films and single strands of polythiophene using atomic force microscopy. *Nat. Commun.* **10**, 1537 (2019).
7. Payam, A. F. & Passian, A. Imaging beyond the surface region: Probing hidden materials via atomic force microscopy. *Sci. Adv.* **9**, eadg8292 (2023).
8. Chiodini, S., Kerfoot, J., Venturi, G., Mignuzzi, S., Alexeev, E. M., Rosa, B. T., Tongay, S., Taniguchi, T., Watanabe, K., Ferrari, A. C. & Ambrosio, A. Moiré Modulation of Van Der Waals Potential in Twisted Hexagonal Boron Nitride. *ACS Nano* **16**, 7589-7604 (2022).

9. Nievergelt, A. P., Banterle, N., Andany, S. H., Gonczy, P. & Fantner, G. E. High-speed photothermal off-resonance atomic force microscopy reveals assembly routes of centriolar scaffold protein SAS-6. *Nat. Nanotechnol.* **13**, 696-701 (2018).

10. Nievergelt, A. P., Erickson, B. W., Hosseini, N., Adams, J. D. & Fantner, G. E. Studying biological membranes with extended range high-speed atomic force microscopy. *Sci. Rep.* **5**, 11987 (2015).

11. Datta, S., Kato, Y., Higashiharaguchi, S., Aratsu, K., Isobe, A., Saito, T., Prabhu, D. D., Kitamoto, Y., Hollamby, M. J., Smith, A. J., Dagleish, R., Mahmoudi, N., Pesce, L., Perego, C., Pavan, G. M. & Yagai, S. Self-assembled poly-catenanes from supramolecular toroidal building blocks. *Nature* **583**, 400-405 (2020).

12. Marchesi, A., Umeda, K., Komekawa, T., Matsubara, T., Flechsig, H., Ando, T., Watanabe, S., Kodera, N. & Franz, C. M. An ultra-wide scanner for large-area high-speed atomic force microscopy with megapixel resolution. *Sci. Rep.* **11**, 13003 (2021).

13. Ando, T., Uchihashi, T. & Fukuma, T. High-speed atomic force microscopy for nano-visualization of dynamic biomolecular processes. *Prog. Surf. Sci.* **83**, 337-437 (2008).

14. Ando, T., Uchihashi, T. & Scheuring, S. Filming Biomolecular Processes by High-Speed Atomic Force Microscopy. *Chem. Rev.* **114**, 3120-3188 (2014).

15. Homma, H., Ngo, K. X., Yoshioka, Y., Tanaka, H., Inotsume, M., Fujita, K., Ando, T. &

Okazawa, H. PQBP3/NOL7 is an intrinsically disordered protein. *Biochem. Biophys. Res. Commun.* **736**, 150453 (2024).

16. Matsushima, K., Sumikama, T., Suzuki, T., Ito, M., Nagasawa, Y., Sumino, A., Flechsig, H., Ogoshi, T., Umeda, K., Kodera, N., Murakoshi, H. & Shibata, M. Structural dynamics of mixed-subunit CaMKIIα/β heterododecamers filmed by high-speed AFM. *Nat. Commun.* **16**, 10603 (2025).

17. Akter, L., Flechsig, H., Marchesi, A. & Franz, C. M. Observing Dynamic Conformational Changes within the Coiled-Coil Domain of Different Laminin Isoforms Using High-Speed Atomic Force Microscopy. *Int. J. Mol. Sci.* **25**, 1951 (2024).

18. Nguyen, P. D. N., Abe, H., Ono, S. & Kodera, N. Actin monomers influence the interaction between Xenopus cyclase-associated protein 1 and actin filaments. *bioRxiv*, 2025.2008.2014.670363 (2025).

19. Umeda, K., Kurokawa, Y., Murayama, Y. & Kodera, N. Submolecular video-imaging of the Smc5/6 complex topologically bound to DNA. *bioRxiv*, 2025.2008.2015.670630 (2025).

20. Muramatsu, D., Watanabe-Nakayama, T., Tsuji, M., Umeda, K., Hikishima, S., Nakano, H., Sakashita, Y., Ikeda, T., Konno, H., Kodera, N., Ando, T., Noguchi-Shinohara, M. & Ono, K. ALZ-801 prevents amyloid β-protein assembly and reduces cytotoxicity: A preclinical experimental study. *FASEB J.* **39**, e70382 (2025).

21. Biyani, M., Isogai, Y., Sharma, K., Maeda, S., Akashi, H., Sugai, Y., Nakano, M., Kodera, N.,

Biyani, M. & Nakajima, M. High-speed atomic force microscopy and 3D modeling reveal the structural dynamics of ADAR1 complexes. *Nat. Commun.* **16**, 4757 (2025).

22. Neaz, M. S., Shoukat, N., Kodera, N. & Sato, H. HS-AFM visualizes stepwise condensation dynamics in RNA-driven LLPS. *bioRxiv*, 2025.2012.2008.693082 (2025).

23. Suzuki, T., Sumikama, T., Matsushima, K., Hasegawa, K., Sumino, A., Umeda, K., Kodera, N., Das, T., Beta, C., Murakoshi, H. & Shibata, M. CaMKIIα holoenzymes self-organize into worm-like mesoscale clusters. *bioRxiv*, 2026.2001.2021.700754 (2026).

24. Ikeda, T., Nojima, T., Yamamoto, S., Yamada, R., Niwa, T., Konno, H. & Taguchi, H. Seesaw protein: Design of a protein that adopts interconvertible alternative functional conformations and its dynamics. *Proc. Natl. Acad. Sci. USA* **122**, e2412117122 (2025).

25. Preiner, J., Kodera, N., Tang, J. L., Ebner, A., Brameshuber, M., Blaas, D., Gelbmann, N., Gruber, H. J., Ando, T. & Hinterdorfer, P. IgGs are made for walking on bacterial and viral surfaces. *Nat. Commun.* **5**, 4394 (2014).

26. Miyagi, A., Chipot, C., Rangl, M. & Scheuring, S. High-speed atomic force microscopy shows that annexin V stabilizes membranes on the second timescale. *Nat. Nanotechnol.* **11**, 783-790 (2016).

27. Imai, H., Uchiumi, T. & Kodera, N. Direct visualization of translational GTPase factor pool formed around the archaeal ribosomal P-stalk by high-speed AFM. *Proc. Natl. Acad. Sci. USA* **117**, 32386-32394 (2020).

28. Hayakawa, Y., Takaine, M., Ngo, K. X., Imai, T., Yamada, M. D., Behjat, A. B., Umeda, K., Hirose, K., Yurtsever, A., Kodera, N., Tokuraku, K., Numata, O., Fukuma, T., Ando, T., Nakano, K. & Uyeda, T. Q. Actin-binding domain of Rng2 sparsely bound on F-actin strongly inhibits actin movement on myosin II. *Life Sci. Alliance* **6**, e202201469 (2022).

29. Ngo, K. X., Vu, H. T., Umeda, K., Trinh, M. N., Kodera, N. & Uyeda, T. Deciphering the actin structure-dependent preferential cooperative binding of cofilin. *eLife* **13**, RP95257 (2024).

30. Kodan, A., Amyot, R., Umeda, K., Ogasawara, F., Kimura, Y., Kodera, N. & Ueda, K. Direct Visualization of ATP-Binding Cassette Protein A1 Mediated Nascent High-Density Lipoprotein Biogenesis by High-Speed Atomic Force Microscopy. *Nano Lett.* **25**, 13563-13570 (2025).

31. Rupitsch, S. J. *Piezoelectric Sensors and Actuators: Fundamentals and Applications*. (Springer Berlin, Heidelberg, 2019).

32. Yong, Y. K., Moheimani, S. O. R., Kenton, B. J. & Leang, K. K. Invited Review Article: High-speed flexure-guided nanopositioning: Mechanical design and control issues. *Rev. Sci. Instrum.* **83**, 121101 (2012).

33. Yong, Y. K. Preloading Piezoelectric Stack Actuators in High-Speed Nanopositioning Systems. *Front. Mech. Eng.* **2**, 8 (2016).

34. Yothers, M. P., Browder, A. E. & Bumm, L. A. Real-space post-processing correction of

thermal drift and piezoelectric actuator nonlinearities in scanning tunneling microscope images. *Rev. Sci. Instrum.* **88**, 013708 (2017).

35. Hosseini, N., Nievergelt, A. P., Adams, J. D., Stavrov, V. T. & Fantner, G. E. A monolithic MEMS position sensor for closed-loop high-speed atomic force microscopy. *Nanotechnology* **27**, 135705 (2016).

36. Tian, Y. L., Ma, Y., Lu, K. K., Yang, M. X., Zhao, X. L., Wang, F. J. & Zhang, D. W. Modeling and control methodology for an XYZ micro manipulator. *Rev. Sci. Instrum.* **90**, 105007 (2019).

37. Tao, Y. D., Li, H. X. & Zhu, L. M. Hysteresis modeling with frequency-separation-based Gaussian process and its application to sinusoidal scanning for fast imaging of atomic force microscope. *Sensor. Actuat. A-Phys.* **311**, 112070 (2020).

38. Fleming, A. J. & Moheimani, S. O. R. Sensorless vibration suppression and scan compensation for piezoelectric tube nanopositioners. *IEEE Trans. Control Syst. Technol.* **14**, 33-44 (2006).

39. Kageshima, M., Togo, S., Li, Y. J., Naitoh, Y. & Sugawara, Y. Wideband and hysteresis-free regulation of piezoelectric actuator based on induced current for high-speed scanning probe microscopy. *Rev. Sci. Instrum.* **77**, 103701 (2006).

40. Fleming, A. J. & Leang, K. K. Charge drives for scanning probe microscope positioning stages. *Ultramicroscopy* **108**, 1551-1557 (2008).

41. Moheimani, S. O. R. & Yong, Y. K. Simultaneous sensing and actuation with a piezoelectric tube scanner. *Rev. Sci. Instrum.* **79**, 073702 (2008).

42. Kos, T., Rojac, T., Petrovcic, J. & Vrancic, D. Control system for automated drift compensation of the stand-alone charge amplifier used for low-frequency measurement. *AIP Adv.* **9**, 035133 (2019).

43. Song, D. W. & Li, C. J. Modeling of piezo actuator's nonlinear and frequency dependent dynamics. *Mechatronics* **9**, 391-410 (1999).

44. Dirscherl, K., Garnæs, J., Nielsen, L., Jøgensen, J. F. & Sorensen, M. P. Modeling the hysteresis of a scanning probe microscope. *J. Vac. Sci. Technol. B* **18**, 621-625 (2000).

45. Croft, D., Shed, G. & Devasia, S. Creep, hysteresis, and vibration compensation for piezoactuators: Atomic force microscopy application. *J. Dyn. Syst. Meas. Control* **123**, 35-43 (2001).

46. Clayton, G. M., Tien, S., Leang, K. K., Zou, Q. Z. & Devasia, S. A Review of Feedforward Control Approaches in Nanopositioning for High-Speed SPM. *J. Dyn. Syst. Meas. Control* **131**, 061101 (2009).

47. Leang, K. K., Zou, Q. Z. & Devasia, S. Feedforward Control of Piezoactuators in Atomic Force Microscope Systems. *IEEE Control Syst. Mag.* **29**, 70-82 (2009).

48. Lu, H., Fang, Y. C., Ren, X. & Zhang, X. B. Improved direct inverse tracking control of a piezoelectric tube scanner for high-speed AFM imaging. *Mechatronics* **31**, 189-195 (2015).

49. Manual, V. (Accessed December 1, 2025).

50. Manual, A. (Accessed December 1, 2025).

51. Fu, J., Chu, W., Dixson, R., Orji, G. & Vorburger, T. Correction of Hysteresis in SPM Images by a Moving Window Correlation Method. *AIP Conf. Proc.* **1173**, 280 (2009).

52. Zhang, L. S., Chen, X. B., Huang, J. C., Li, H. L., Chen, L. J. & Huang, Q. X. A method to correct hysteresis of scanning probe microscope images based on a sinusoidal model. *Rev. Sci. Instrum.* **90**, 023704 (2019).

53. Dickbreder, T., Sabath, F., Höltkemeier, L., Bechstein, R. & Kühnle, A. unDrift: A versatile software for fast offline SPM image drift correction. *Beilstein J. Nanotech.* **14**, 1225-1237 (2023).

54. Kubo, S., Umeda, K., Kodera, N. & Takada, S. Removing the parachuting artifact using two-way scanning data high-speed atomic force microscopy. *Biophys. Physicobiology* **20**, e200006 (2023).

55. Fukuda, S. & Ando, T. Faster high-speed atomic force microscopy for imaging of biomolecular processes. *Rev. Sci. Instrum.* **92**, 033705 (2021).

56. Bain, A. K. & Chand, P. *Ferroelectrics: Principles and Applications*. (Wiley‐VCH Verlag GmbH & Co. KGaA, 2017).

57. Meier, D. & Selbach, S. M. Ferroelectric domain walls for nanotechnology. *Nat. Rev. Mater.* **7**, 157-173 (2022).

58. Oling, F., Bergsma-Schutter, W. & Brisson, A. Trimers, dimers of trimers, and trimers of trimers are common building blocks of annexin A5 two-dimensional crystals. *J. Struct. Biol.* **133**, 55-63 (2001).

59. Jung, H. & Gweon, D. G. Creep characteristics of piezoelectric actuators. *Rev. Sci. Instrum.* **71**, 1896-1900 (2000).

60. Gu, G. Y., Zhu, L. M., Su, C. Y., Ding, H. & Fatikow, S. Modeling and Control of Piezo-Actuated Nanopositioning Stages: A Survey. *IEEE Trans. Autom. Sci. Eng.* **13**, 313-332 (2016).

61. Hao, G. L., Cao, K. R., Li, R., Li, Z. K., Du, H. R. & Tan, L. Y. Rate-dependent hysteresis modeling and compensation for fast steering mirrors. *Sensor. Actuat. A-Phys.* **376**, 115568 (2024).

62. Horcas, I., Fernandez, R., Gomez-Rodriguez, J. M., Colchero, J., Gomez-Herrero, J. & Baro, A. M. WSXM: A software for scanning probe microscopy and a tool for nanotechnology. *Rev. Sci. Instrum.* **78**, 013705 (2007).

63. Kizu, R., Misumi, I., Hirai, A., Kinoshita, K. & Gonda, S. Development of a metrological atomic force microscope with a tip-tilting mechanism for 3D nanometrology. *Meas. Sci. Technol.* **29**, 075005 (2018).